\documentclass[conference]{IEEEtran}
\ifCLASSINFOpdf
   \usepackage[pdftex]{graphicx}
   \DeclareGraphicsExtensions{.pdf,.jpeg,.png}
\else
\fi

\usepackage{algorithm}

\usepackage{algpseudocode}
\usepackage{siunitx}
\usepackage{subfig}
\usepackage{cite}
\usepackage[final]{changes}

\newcommand{\enesays}[1]{\color{red} #1 \color{black}}
\renewcommand{\enesays}[1]{}

\definechangesauthor[name={Dominik}, color=red]{dominik}
\definechangesauthor[name={Suat}, color=blue]{suat}
\definechangesauthor[name={Mumin}, color=purple]{mumin}
\definechangesauthor[name={Enes}, color=purple]{enes}
\usepackage{atbegshi,picture}
\usepackage{lipsum}

\AtBeginShipout{\AtBeginShipoutUpperLeft{%
  \put(\dimexpr\paperwidth-1cm\relax,-1.5cm){\makebox[0pt][r]{\framebox{Part of this paper will be presented in IEEE ICBC, 2020, Toronto}}}%
}}

\begin{document}
\linespread{1}
%

\title{Improving Transaction Success Rate via Smart Gateway Selection in Cryptocurrency Payment Channel Networks}

\author{\IEEEauthorblockN{Suat Mercan, Enes Erdin and Kemal Akkaya} 
\IEEEauthorblockA{Dept. of Electrical and Computer Engineering\\ Florida International University, Miami, FL 33174\\ Email: \{smercan, eerdi001, kakkaya\}@fiu.edu}
}

\maketitle

\begin{abstract}

The last decade has experienced a vast interest in Blockchain-based cryptocurrencies with a specific focus on the applications of this technology. However, slow confirmation times of transactions and unforeseeable high fees hamper their wide adoption for micro-payments. The idea of establishing payment channel networks is one of the many proposed solutions to address this scalability issue where nodes, by utilizing smart contracting, establish payment channels between each other and perform off-chain transactions. However, due to the way these channels are created, both sides have a certain one-way capacity for making transactions. Consequently, if one sides exceeds this one-way capacity, the channel becomes useless in that particular direction, which causes failures of payments and eventually creates an imbalance in the overall network. 
To keep the payment channel network sustainable, in this paper, we aim to increase the overall success rate of payments 
by effectively exploiting the fact that end-users are usually connected to the network at multiple points (i.e., gateways) any of which can be used to initiate the payment. We propose an efficient method for selection of the gateway for a user by considering the gateway's inbound and outbound payment traffic ratio. We then augment this proposed method with split payment capability to further increase success rate especially for large transactions. The evaluation of the proposed method shows that compared to  greedy and maxflow-based approaches, we can achieve much higher success rates, which are further improved with split payments. 

\end{abstract}

%
\maketitle

\section{Introduction}

Cryptocurrency, a path-breaking invention for the storage and transfer of assets, is one of the most influential technologies of 2010s \cite{cryptocurrency}. The intriguing idea behind it is storing the ownership of the entities in an append-only, tamper-proof database which is known as Blockchain.  
Although Blockchain is being used in various applications \cite{blockchainapp}, Bitcoin is the first practical and widely accepted use case. 
Consensus mechanism (i.e., Proof of Work (PoW)) used in Blockchains eliminates the necessity of a central authority to approve and keep the records. This enables making transactions in a trustless environment. Blockchain is also irreversible and publicly available, so all the participants have the copy of the data, thus, the system is resilient against node failures and malicious acts. Therefore, it also solves the single point of failure problem and removes dependency on the trusted third parties for transactions. 

Even though the concept of a virtual currency is a brilliant idea with a promising architecture, it still suffers from lack of wide adoption due to its impracticality in day-to-day micro-payments stemming from two main factors; 
1) high transaction fees and; 2) long block confirmation times.  For instance, in Bitcoin it takes 10 minutes to approve a block of transactions. Furthermore, to prevent double spending, as a rule of thumb, the merchants wait for approval of 6 consecutive blocks. Additionally, the size of a block is limited to 4 MB which not only hinders the possibility of an approval of a payment on a congested day but also limits the total number of transactions in a unit time. The theoretical maximum in Bitcoin is calculated to be 7 transactions per second\cite{tranrate} which is far lower than what Visa or MasterCard can process\cite{visatran}. The energy spent by the miners is another factor in the valuation of the transaction fees \cite{bitcoinfee}.

There has been some recent attempts to address these issues. Payment channel network (also known as off-chain transaction networks) concept \cite{ln,raiden} is among the proposed solutions. The idea is to utilize smart-contracts and avoid writing every transaction on the blockchain. Instead, the transactions are recorded \textit{off-chain} until the accounts are reconciled. Specifically, once a channel is created between two peers, many transactions can be performed in both directions as long as there is enough funds. When many nodes come together, the off-chain transaction channels turn into a network of payment channels. Instead of opening a direct channel, a peer makes use of an already established channel to forward money over existing nodes by paying a \textit{transaction fee} as long as a path exists from the payer to the payee. This also helps in reducing  channel opening costs for a user since existing channels can be utilized.  
Lightning Network (LN) is a perfect example of this concept that has reached to almost 10K users in 2 years \cite{ln} for Bitcoin transactions. This payment network has specific nodes which charge transaction fees to users for passing their data over these nodes. Basically, a user connects to a \textit{gateway node}, which is further connected to \textit{routing nodes} that act as the network backbone. Ultimately, LN achieves almost real-time transactions with negligible fees compared to Bitcoin fees and transaction validation times.


Nevertheless, payment channel networks, including LN, come with their own challenges due to the way the channel capacities are consumed. Basically, if there is not enough available capacity on a channel, the transactions cannot be sent. More specifically, the channels are bi-directional and when they are created, each peer's (say $A$ and $B$) one-way capacity is set independently. As a result, a channel's capacity in one direction (i.e., from $A$ to $B$) can be totally consumed while the capacity in the opposite direction (i.e., from $B$ to $A$) holds all of the funds. This means, while one of the peers, $B$, can make transactions in one direction, the other peer, $A$, cannot make any transactions through the channel. To be able to send transactions again, $A$ needs to receive payments from $B$ so that it can increase its channel capacity. Due to this feature, insufficient funds in one-way channels drastically drops the chance of payment transfers in the network. One of the recent studies performed on LN \cite{ln}, the most prominent implementation of this concept, indicates that chance of sending a \$5 payment successfully is around 50\% which makes it practically useless for end-users. Therefore, there is a need to address this issue to be able to increase the success rate of transactions in payment channel networks. To this end, we propose modifying the payment route selection in payment channel networks. 



Before explaining the rationale of our approach, we would like to describe current LN's approach to routing of payments. LN advocates a self management and free market model. Node operators are responsible to set the fee, select the connections, and find the path to send the payment. Thus, transaction fee is one of the major factors for end-users when choosing a route. Therefore, the users tend to route their payments through the cheapest path. We refer to this approach as \textit{greedy} method since it focuses on minimum fee and represents a selfish approach. Obviously, this approach will not help in balancing the channels since their capacities should be accounted when finding the paths. Therefore, a potential solution would be to compute maximum flow from a source to destination and leaving the fee as a secondary factor. However, this flow maximization needs to consider the whole network to come up with a more balanced approach. 

Therefore, in this paper, we propose a new routing approach named \textit{gateway inbound/outbound ratio} which calculates the ratio of total inbound capacity to outbound capacity of each connected gateway and chooses the minimum among these. The insight that leads us to this method is that the bottleneck in payment channel networks is the gateways where the payments are initiated from or destined to the users. Routing nodes that constitute the backbone of the payment network are naturally being balanced since they transmit high number of transactions in both ways. However, gateway nodes need to be taken care of explicitly. 

To further increase the success rate, we then focuse on split payments, where a payment can be divided into pieces and sent independently so that channel capacity variations will not be major. We looked into two different split approaches; \textit{equal share} which distributes the total amount among the gateways, and \textit{proportional split} which assigns the payments to gateways based on their inbound outbound ratios.

We implemented and tested the effectiveness of the proposed approaches under various payment scenarios and observed that the success rate can be significantly improved with proper gateway selection and proportional split payment. It creates symmetrically balanced channels and improve the efficiency of the payment channel networks.

This paper is organized as follows: Section \ref{sec:Related} summarizes the related work in the literature while Section \ref{sec:Preliminary} provides some background explaining the concepts used in payment networks and our assumptions. Section \ref{sec:approach} presents the problem definition and our approach. In Section V, we assess the performance of the proposed mechanism. Finally, Section VI concludes the paper.

\section{Related Work}
\label{sec:Related}

Various methods for the realization of the payment channel network idea have been suggested and some are already being implemented. LN \cite{ln} for Bitcoin and Raiden \cite{raiden} for Ethereum are two examples in practice. LN is the most active one with more than 10,000 nodes and 30,000 channels \cite{acinq}. LN utilizes source-routing for transferring payments. The node first discovers a path with available channel capacity, then initiates the transaction. Spider \cite{spider} applies packet-switching routing techniques to payment channel networks. The payments are split into micro-payments similar to maximum transmission unit (MTU) in computer networks. It uses congestion control and a best-effort model to improve payment throughput. It specifically chooses the paths that re-balances the channels. The payments are queued at spider routers and they are transferred when the fund is available. Flash \cite{flash} applies a distributed routing algorithm to better handle constantly changing balances. It differentiates mice and elephant payments. Small payments are sent randomly over pre-computed paths. For large payments, it probes the nodes to find the channel with available funds. Then it splits the payment into several smaller chunks. Revive \cite{revive} assumes that a node has multiple connections and the skewed payments make some of the links depleted. It tries to find cycles in the network and a user sends a payment to herself to re-balance the depleted channel through others. SpeedyMurmur \cite{speedymurmur} is focusing on the privacy of the payments by using an embedding-based routing algorithm. SilentWhisper \cite{silentwhisper} and Flare \cite{flare} are using landmark routing in which only some nodes store routing tables for the complete network. The rest only knows how to reach to one of those landmark nodes. A user transmits the payment to the gateway node which handles the rest. Our work in this paper focuses on path selection strategies to choose the appropriate route for payments so that the overall network will be more sustainable (i.e., more successful transactions). In this regard, we explore the gateway selection strategy which has a great impact on the overall success of payment channel networks. This topic was not well in those works as they mostly focus on routing mechanisms.

\section{Preliminaries and Assumptions}
\label{sec:Preliminary}

\subsection{Blockchain and Bitcoin} 
Blockchain is a distributed database in which the building data structure is called a ``block''. For Bitcoin, a block is simply comprised of transactions (data), timestamp, nonce, the hash of the block and the hash of the previous block\cite{bitcoin}. 
In cryptocurrency-based networks, the nodes come to a consensus for the approval of a block by proving that they have enough interest in the network. For instance in Bitcoin, Hashcash proof-of-work (PoW) mechanism is utilized. In order for a block to be accepted as valid, the hash of the block should be smaller than a number which is decided by considering the total accumulated computational power in the entire network. 
By changing the nonce value in the block the miner aims to find a suitable hash result. Soon after a valid block is found, it gets distributed in the network. After the other nodes validate that block, the next block calculation starts.

\subsection{Off-Chain Payment Channels}
For Bitcoin, the average time spent for the approval of a block is 10 minutes. This duration casts suspicion on the usability and practicality of the Bitcoin. More precisely, using Bitcoin for day-to-day spending becomes almost impossible. The reason for that is a payee waits at least 6 blocks to count a transaction to be valid. So, for example, if one buys a cup of coffee and uses Bitcoin to pay, s/he has to wait at least one hour for the payment to get cleared. Moreover, during the congested times, s/he has to either pay a lot of transaction fees, possibly greater than the price of the coffee or has to wait much more than the anticipated time. Apparently, this is an undesirable case not only for a customer but also for a shop owner. 

To solve that problem, developers came up with the concept of a mechanism called ``off-chain payment channel'' that leverages the smart contract mechanism in blockchain. In that concept, two users, say A and B, come to a mutual agreement on establishing a business. Then they sign a contract by transferring collateral to a shared 2-of-2 multi-signature address and initiate the channel by publishing it on the blockchain. This contract type is called ``Hash Time Locked Contracts'' (HTLC). When the users agree on any amount of payment, they prepare a new HTLC, exchange the new contract, and update the state of the channel. To initiate a payment from a payer, a challenge, namely, a pre-image is sent to the recipient. If the recipient can reply successfully to the challenge, the contract becomes valid, and the ownership of the money gets transferred. Off-chain mechanism brings a huge advantage since the peers do not need to publish every transaction on the blockchain. That is, the payments are theoretically instantaneous. Moreover, as there is no need for frequent on-chain transactions, the transactions will be protected from fluctuating, unexpectedly high on-chain transaction fees. In fact, a transaction fee can be 0 (zero) if the peers agree so. Thus, for a well-defined channel between honest peers, there will be two on-chain transactions: one to establish the channel and one to close (finalize) the channel.

\begin{figure}[htb]
    \centering
    \includegraphics[width=0.7\linewidth]{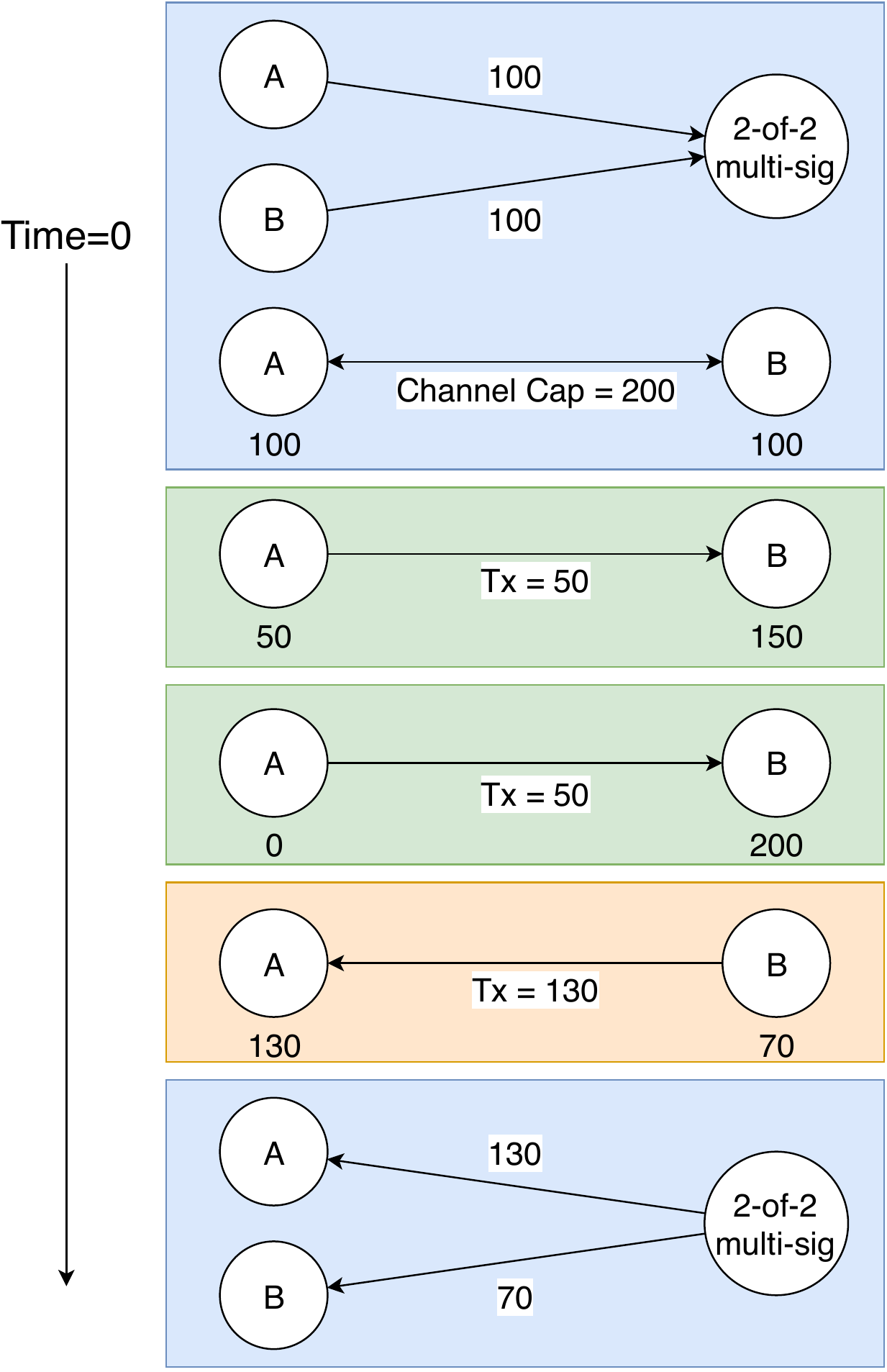}
    \caption{Illustration of a payment channel.}
    \label{fig:channels}
\end{figure}

An important feature of such a channel is that the direction of the payments matter. Specifically, two flows from opposite directions on the same link negate each other's capacity consumption. This is shown in Fig. \ref{fig:channels}. At Time=0, a channel between two parties $A$ and $B$ is established. Both $A$ and $B$ put 100 unit of currency which in turn makes the channel capacity 200 units. After $A$ makes 2 transactions each of which is 50 units, the directional capacity from $A$ to $B$ will be zero. Hence, $A$ can not transfer any more unless $B$ transfers back some money. $B$ sends 130 unit back and after they close the channel they get their corresponding shares from the multi-signature address.

\subsection{Payment Channel Networks}
\begin{figure}[htb]
    \centering
    \includegraphics[width=0.9\linewidth]{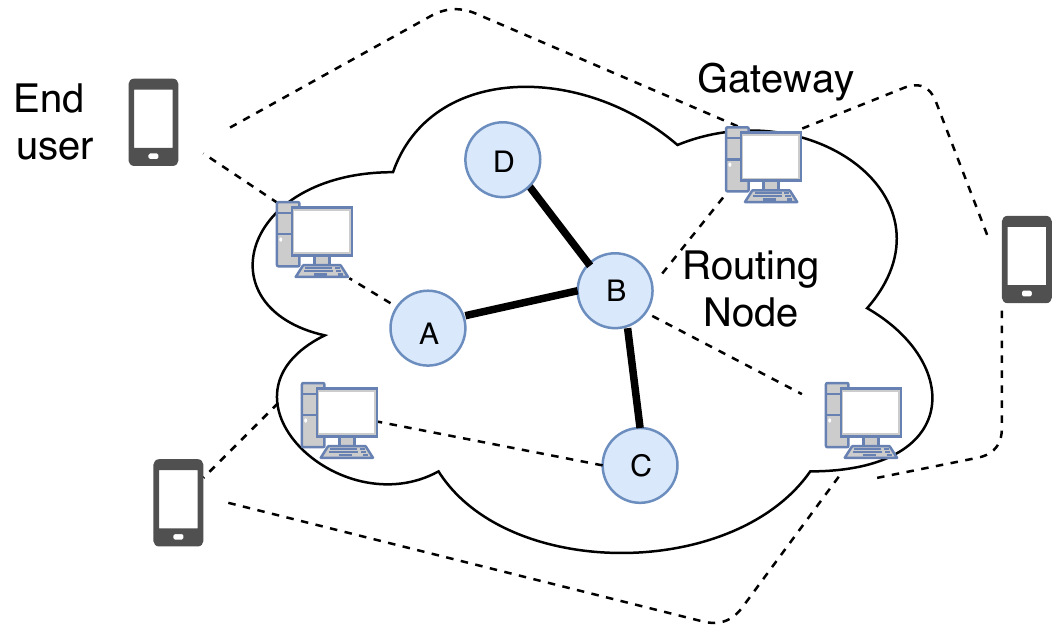}
    \caption{Payment channel network.}
    \label{fig:PBFT}
\end{figure}

Off-chain payment channels can be extended to a payment channel network idea. As shown in Fig.~\ref{fig:PBFT}, assume that $A$ and $B$ have a channel, and $B$ and $C$ have a channel too. If somehow, $A$ wants to trade with $C$ only, what s/he has to do is hash-lock a certain amount of money and forward it to $C$ through $B$. As $C$ already knows the answer to the challenge, $C$ will get her/his money from $B$ by disclosing the answer. The brilliance of the HTLC appears here. As $C$ discloses the answer to the challenge, $B$ learns the answer. Now, $B$ will reply to the challenge successfully and get her/his share from $A$. In this way, one can reach everyone in a network through multi-hop payments forming a payment channel network. 

A payment channel network illustrated in Fig.~\ref{fig:PBFT} consists of three types of nodes in terms of functionality; end-user, gateway, and routing nodes: 
\begin{itemize}
\item \textit{End-user}: An end-user usually makes payment to purchase an item and rarely receives payment such as refund. S/He connects to a gateway to access the rest of the network. S/He needs to maintain certain amount of funds in his/her channel with the gateway to continue sending payments. S/He should open the channel with the best gateway that can connect him to others continuously and cheaply. The best choice would be connecting to a gateway that s/he will usually have direct transactions.

\item \textit{Gateway nodes}: This type of nodes are usually the stores that expect to receive payments in cryptocurrency. They also relay the payments of their end-users' payments to other nodes. Gateway nodes should connect to routing nodes with good connectivity. They may not aim to earn transfer fees. Their primary purpose is to sell their products in cryptocurrency.

\item \textit{Routing nodes}: These nodes typically act as backbone routers (similar to BGP routers on Internet) to connect gateways. They regard having a node in LN as an investment opportunity. They try to have high number of connections and maintain the channels well balanced. This makes them a hub point in the network and lure the transactions so that they increase the return of their investments.
\end{itemize}

\subsection{Assumptions}
In this paper, we consider that a payment channel network consists of nodes connected with off-chain payment channels. The channel capacity represents the amount of money deposited in a 2-of-2 multi-signature address. Although any node can send to and receive payment from any other node, we want to distinguish nodes as end-users, gateways and routing nodes since we want to simulate the case that people want to use cryptocurrency for shopping and micropayment where the cash flow is mostly from a customer to a gateway (a store in real life). After a person establishes a payment channel with a gateway, she can make a payment to anybody in the network.

We build up our work on a presumed existing routing protocol (similar to LN's) which is used by the nodes to advertise the weights to the rest of the network and carry a specified payment from a source to destination. We are not focusing on the efficiency and overhead of the routing protocol as our goal is to determine the gateway to start the payment. Each node is assumed to know the complete topology to calculate the path, and the updates about the channels are propagated to other nodes in the network. 

We would like to note that our approach is designed to be used in any payment channel network. While we use LN as an example to explain some concepts and problems, the proposed approach is not specifically designed to work solely within LN. 


\section{Proposed Approach}
\label{sec:approach}

\subsection{Problem Motivation and Overview}
In a payment channel network, an end-user transfers a payment to a store by initiating the payment via a gateway. This payment first goes through the gateway, then routing nodes and arrives at the destination unless there is a direct channel between the end-user and the gateway. This multi-hop transmission may not be completed because of insufficient funds in the channels. The funds in channels may shift to one particular side of the channels if the payments are highly directional. While routing nodes may not suffer dramatically from this type of problem because of high number of transactions and well balanced channels, the gateway nodes are particularly prone to channel balance exhaustion. 

\begin{figure}[h]
    \centering
    \includegraphics[scale=0.4]{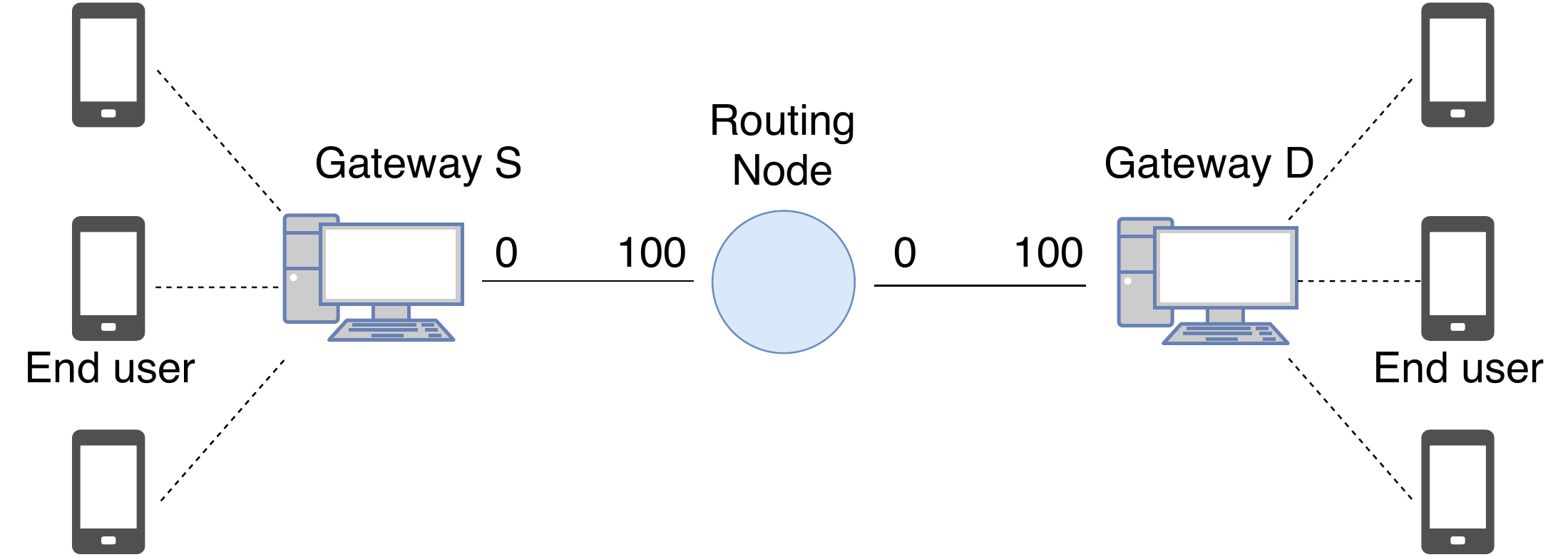}
    \caption{Depleted channels cause disconnection.}
    \label{fig:endnodes}
\end{figure}

\vspace{0.1cm}
\noindent 


This is because, if a gateway has used up its outgoing capacity, then it can not initiate any payment and function as a transit node unless it receives some payment. Suppose that a store is having a busy day and receiving many payments. The channel between the store (gateway) and a routing node will be depleted and the store will not be able to receive payments anymore. This is depicted as an example in Fig.~\ref{fig:endnodes}. In this figure, the end-users connected to Gateway $S$ can not make payments through this node as its outgoing channel capacity is 0. They have to wait for Gateway $S$ to receive some payments destined to it. Similarly, if a gateway consumed all of its incoming capacity, then it can not receive any payment and function as a transit node as well until it is used to send some payment. In Fig.~\ref{fig:endnodes}, Gateway $D$ can not receive any payment since the incoming channel capacity is 0. It has to wait for any end-user to send payments originating from $D$.

Our motivation to address this issue comes from the fact that in payment channel networks, a payment can be initiated from various points. Therefore, this can be an opportunity to balance the channels. Specifically, although we may not control the destinations of the payments, we can influence the initial gateway connection points for users. For example, a node receiving a high volume of payments should also be preferred as the source, and a node which has less outgoing capacity should be avoided to originate payments. In this way, the channels can be re-balanced with the help of cooperative end-user node implementation.

\begin{figure}[!h]
    \centering
    \includegraphics[width=.83\linewidth]{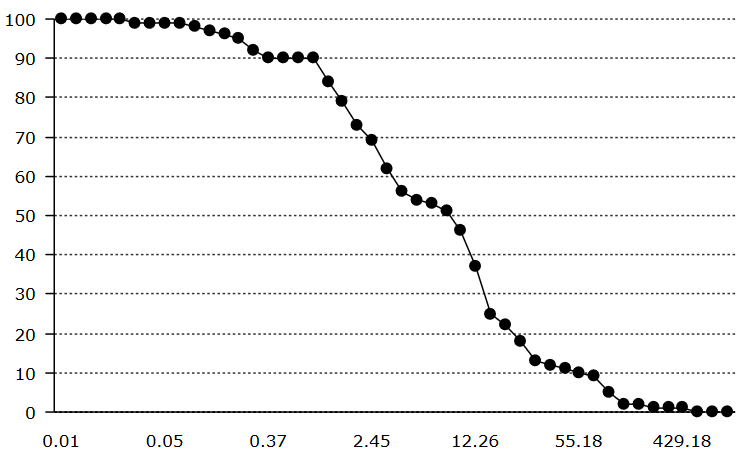}
    \caption{Success rate vs amount of the payment \cite{diar}.}
    \label{fig:diar}
\end{figure}

Another problem with current LN implementation is that it only allows atomic payments, which means that there must be a wide enough single channel to carry a payment from source to destination. In such a system, large payments will cause major shift in channel capacities which might inhibit the success of the transactions in the overall system. The results of the experiment shown in Fig. \ref{fig:diar} performed by \cite{diar} indicates that success rate decreases dramatically with increasing transfer amount. We argue that multiple paths consisting of multiple small channels might better satisfy a larger transaction amount when split payments are allowed.

So in summary, our approach utilizes smart gateway selection and split payments to improve the overall success rate for the transactions in payment channel networks. Next, we detail these ideas. 


\subsection{Smart Gateway Selection}

We propose a gateway selection method that utilizes multiple connections effectively between the end-user and the gateway to create a more stable network. This method focuses on the balances of the gateway channels and strive to keep them balanced both inward and outward. The rationale is balancing incoming and outgoing payments amounts for the gateways so that they are not blocked from sending or receiving future payments. Therefore, we refer to this as \textit{Gateway Inbound/Outbound Ratio} approach. 

The algorithm presented in Alg.\ref{alg:ratio} calculates the total inbound and outbound capacity for each connected gateway, and then it finds the one (gateway) with the lowest ratio. Basically, the inbound/outbound ratio is used as the metric when choosing the gateway to initiate the payment. Since we are using single payment method, the links which do not have enough capacity to transmit the amount is not considered in path computation. If any of the gateway does not have the required capacity, that is ruled out in the path calculation either. When the gateway is decided, the transaction process starts.


\begin{algorithm}
\caption{Inbound/Outbound Ratio}\label{alg:ratio}
\begin{algorithmic}[1]
\State Input: \textit{C=Gateway list}, \textit{G=Connected directed graph}
\For{each directed edge, $(u,v)$, in $G$}
    \State $weight(u,v)$=$bal(v,u)$/[$bal(u,v)$+$bal(v,u)$] 
\EndFor
\hspace{2mm}// \textit{weight calculations are done}
\State $max$ = Integer.Min
\For {every \textit{gateway connection}, $s$ in $C$}
// \textit{Find the gateway with highest inbound/outbound ratio}
    \State $total\_in$ = $total\_out$ = 0
    \For {every \textit{channel}, $(x,y)$, of $gateway$ $s$}
        \State $total\_in$ = $total\_in$  + $capacity(y,x)$
        \State $total\_out$ = $total\_out$  + $capacity(x,y)$
    \EndFor
    \State $ratio$ = $total\_in$ / $total\_out$
    \If{$ratio$ larger than $max$}
        \State $max$ = $ratio$
    \EndIf
\EndFor
\State $Path$=$ShortestPath$($G$, from=$max$, to=$d$) 
\State Output: $Path$
\end{algorithmic}
\end{algorithm}

\subsection{Split Payments}
In order to further increase the success rate for transactions, we also explore various models for splitting the payments. The ability to split a payment might allow big amounts to be transferred by using multiple gateways if a single path is not enough to carry the totality of a transaction. That should increase the success rate and help manage the balances over channels better. 

A payment can be split in many ways, such as dividing into very small units (i.e., micropayments) or equal bigger chunks and these can be split among multiple gateways. Each of these ways can be applied simultaneously along with the smart gateway selection method we described earlier. These methods might incur additional transaction costs depending on the fee policy. For instance, if there is a base fee for each transaction, micropayment model will be infeasible. 
In this paper, we design two different splitting models as detailed below: \\


\noindent \textit{Equal Share}: In this model, the payment request is divided among the gateways in equal shares if they all are capable of transferring that amount. We first check if there is enough total capacity from all connected gateways to the destination using multiple-source single-sink max-flow algorithm \cite{maxalg}. It is modified from max-flow algorithm by adding one extra node to the network which is connected to all gateways with unrestricted channel capacity. If the calculated capacity is not enough, splitting also does not help to perform the transfer because basically there is no route for transferring this amount. 

If we find out that a sufficient capacity exists, then we divide the payment into multiple chunks based on the number of available connections of the end user. We do not consider any criteria of the gateways such as total capacity, fee or ratio. All the gateways will forward equal amounts if possible. However, if any of them does not have enough capacity, then others will have to carry more to compensate for the remaining amounts. The idea is utilizing the channels fairly by distributing the payments among the gateways as much as possible.\\

\noindent \textit{Proportional Split}: 
In this split method, we consider a split criterion based on the gateways' inbound/outband ratios instead of the number of gateways. Basically, we still check if there is enough capacity to carry the payment from source to the destination. However, this time, the total amount is divided proportional to inbound/outbound ratio of the gateways if possible. Specifically, the gateway with less outbound capacity compared to inbound capacity transfers less amount of transactions. This method should help the gateways better to stay balanced especially for the existence of large payments. 

Note that this method is also in line with our gateway selection approach in the previous section as they both consider the same criterion. Thus, we expect that this splitting will further improve the performance in terms of success rate.

\section{Performance Evaluation}
 
This section presents the experiment setup and results for the proposed approach. 

\subsection{Experiment Setup}
We developed a payment channel network simulator in Java that allows us to run the experiments and measure the defined metrics. We list the necessary parameters in Table \ref{table1} to set when running the experiments.

\begin{table}[h]
\renewcommand{\arraystretch}{1.5}
\caption{Experiment Parameters}
\label{table1}
\centering 
\begin{tabular}{|c|c|}
\hline
 Number of Nodes           & 100     \\
\hline 
 Degree of a Node          & 3       \\
\hline 
 Initial Channel Capacities  & 50 to 150  \\
\hline 
 Payment Amount            & 5 to 85    \\
\hline 
 Number of Payment         & 5K      \\
\hline
\end{tabular}
\end{table}

\textit{Network configuration:} The results are based on a random regular network with 100 nodes, each with degree 3. The channels between the nodes are assigned a random capacity uniformly distributed between 50 and 150. 

\textit{Payment files:} We build different payment sequences consisting of 5000 end-to-end transactions. 
Each node sends and receives 50 transactions on average. Each transaction amount is selected randomly during the experiment within a specified range. Source and destination are not necessarily the same which means that node $A$ transmits to $B$ but may receive from $C$. \enesays{what does A transmits to B but may receive from C mean? Also source and destination nodes are not necessarily the same daha uygun gibi. }

\textit{Payment transfer:} Each node calculates the path using specified method and the payment is sent through intermediate nodes by decrementing the amount from each channel used and incrementing in the opposite direction. In the first set of experiment, we use only single payment while in the second part we use split payment.

\textit{Experiment run:} Each payment file is run with various seeds which randomized the initial channel capacity and payment amount. Thus, the results are aggregate of 75 randomized tests on the network.


\begin{figure*}[!ht]

\begin{centering}
\begin{tabular}{ccc}
\includegraphics[keepaspectratio=true,angle=0,width=58mm]{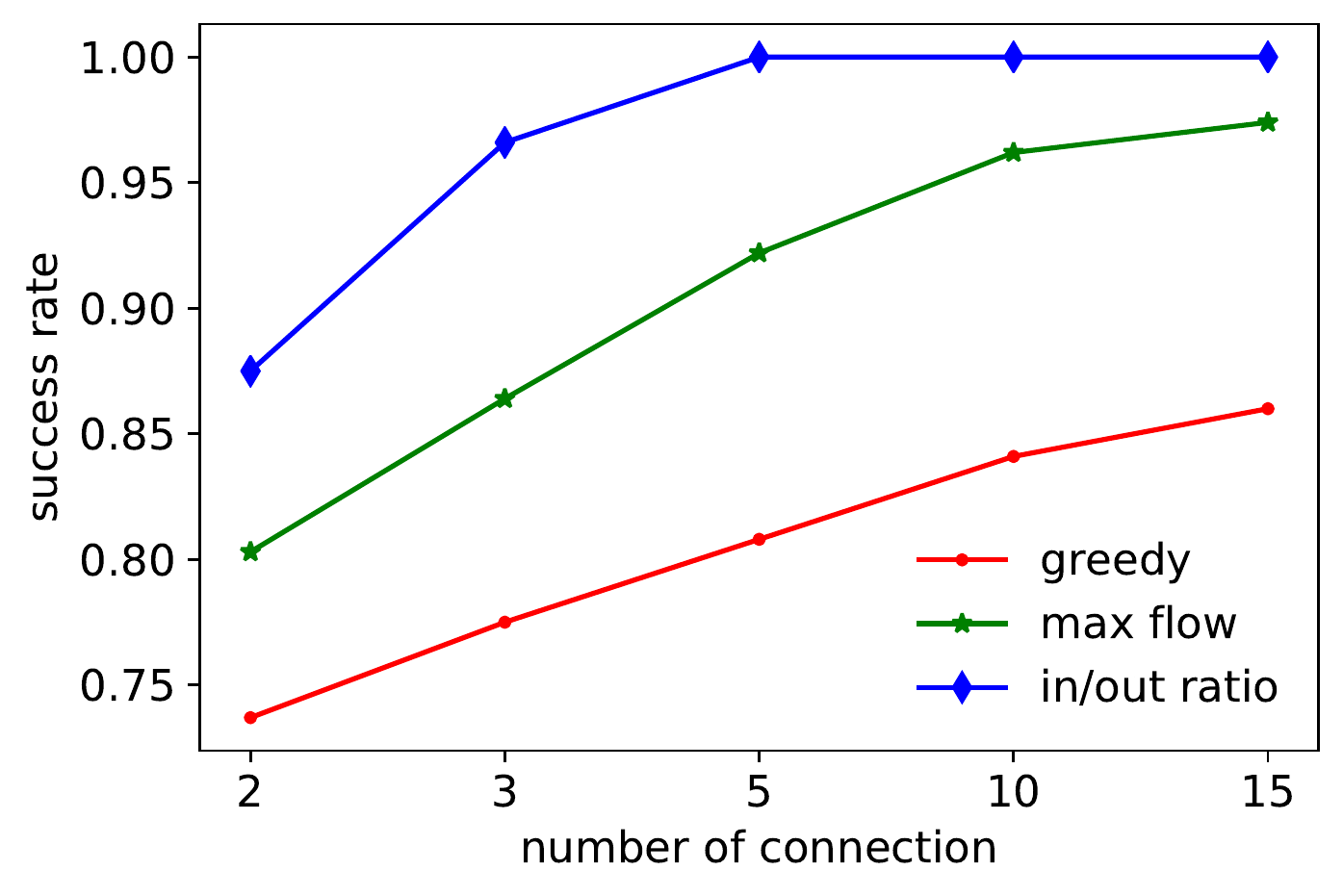} &
\hspace{-5mm}
\includegraphics[keepaspectratio=true,angle=0,width=58mm]{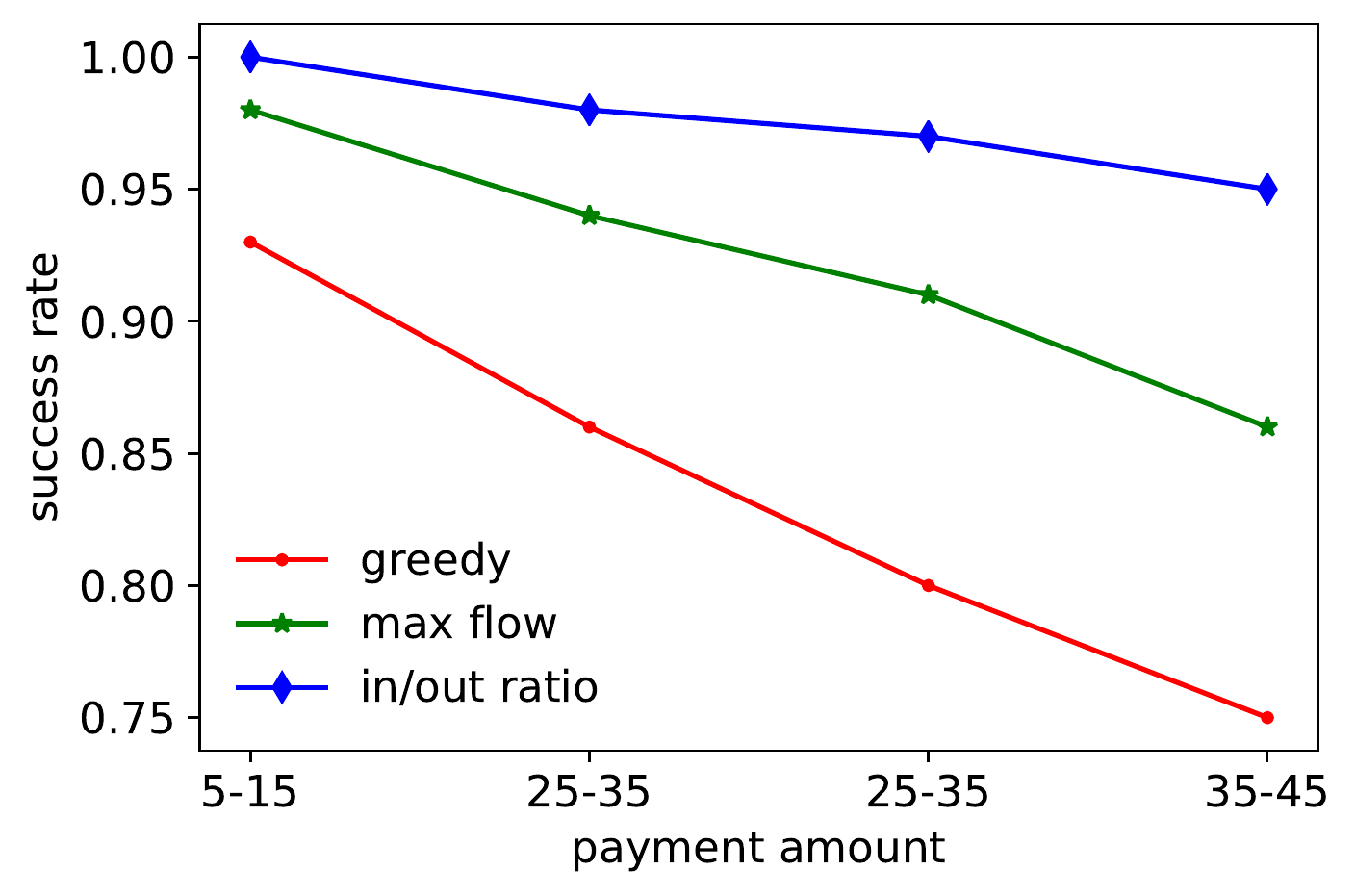} &
\hspace{-5mm}
\includegraphics[keepaspectratio=true,angle=0,width=58mm]{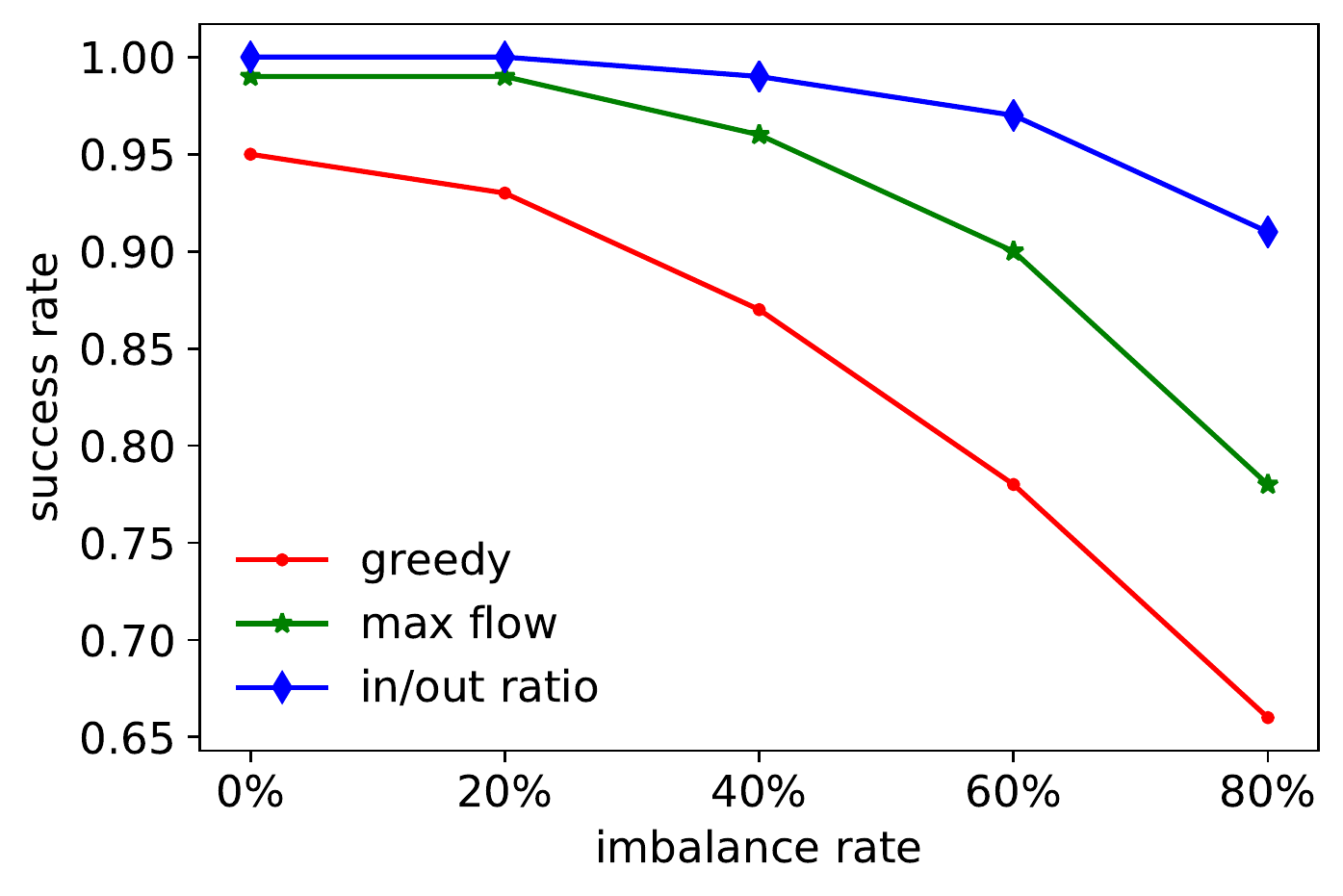} 

\vspace{-2mm}

 \tabularnewline
\footnotesize{ a. Success rate vs. connection count} &
\hspace{-5mm}
\footnotesize{ b. Impact of payment amount } & 
\hspace{-5mm}
\footnotesize{ c. Success rate vs. imbalance rate } 
\vspace{2mm}
 
\tabularnewline

\end{tabular}
\caption{Experiment Results for Success Rate}
\label{fig:res1}
\vspace{-5mm}
\end{centering}
\end{figure*}

\begin{figure*}[!htb]
\begin{centering}
\begin{tabular}{ccc}

\includegraphics[keepaspectratio=true,angle=0,width=58mm]{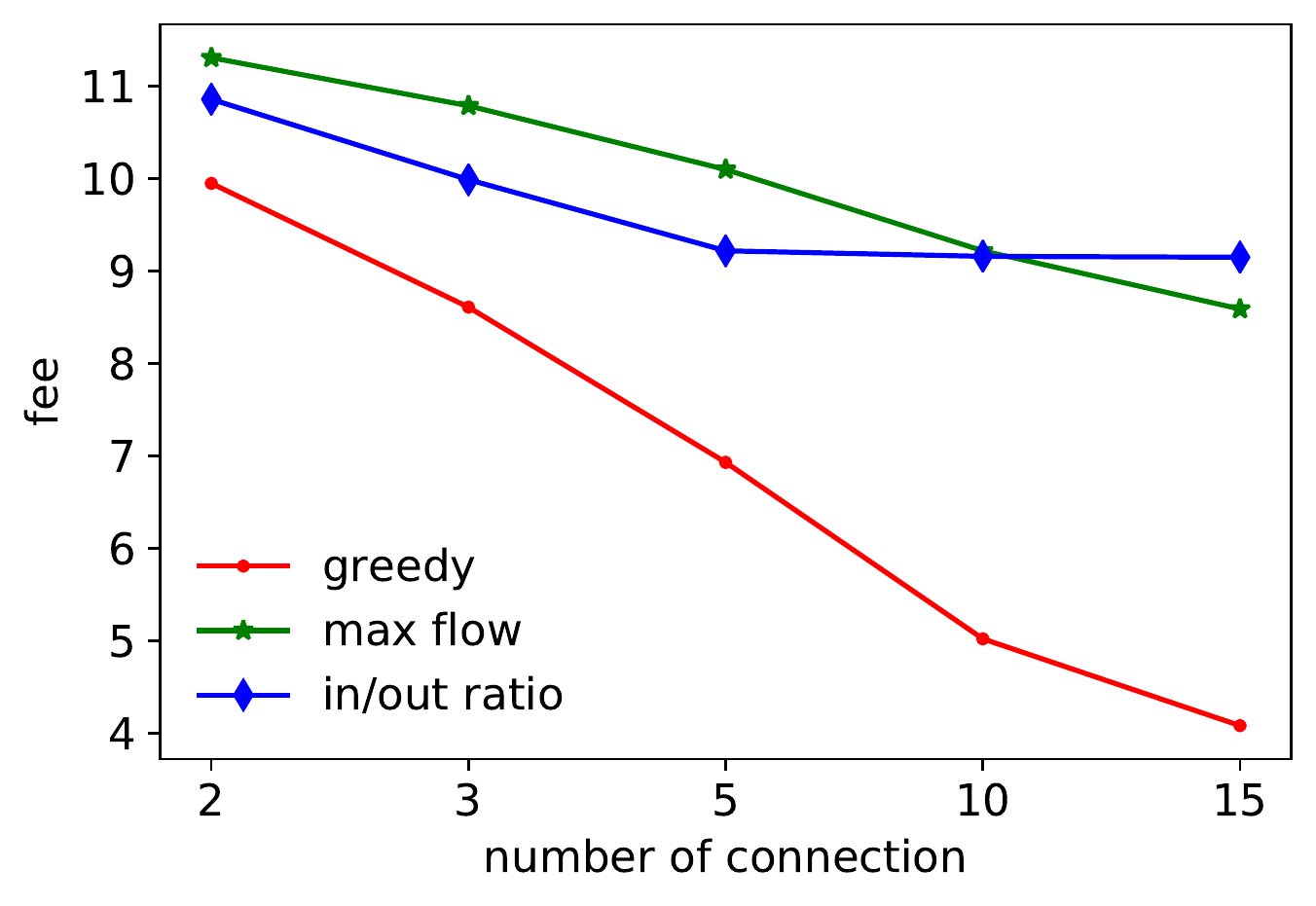} & 
\hspace{-5mm}
\includegraphics[keepaspectratio=true,angle=0,width=57mm]{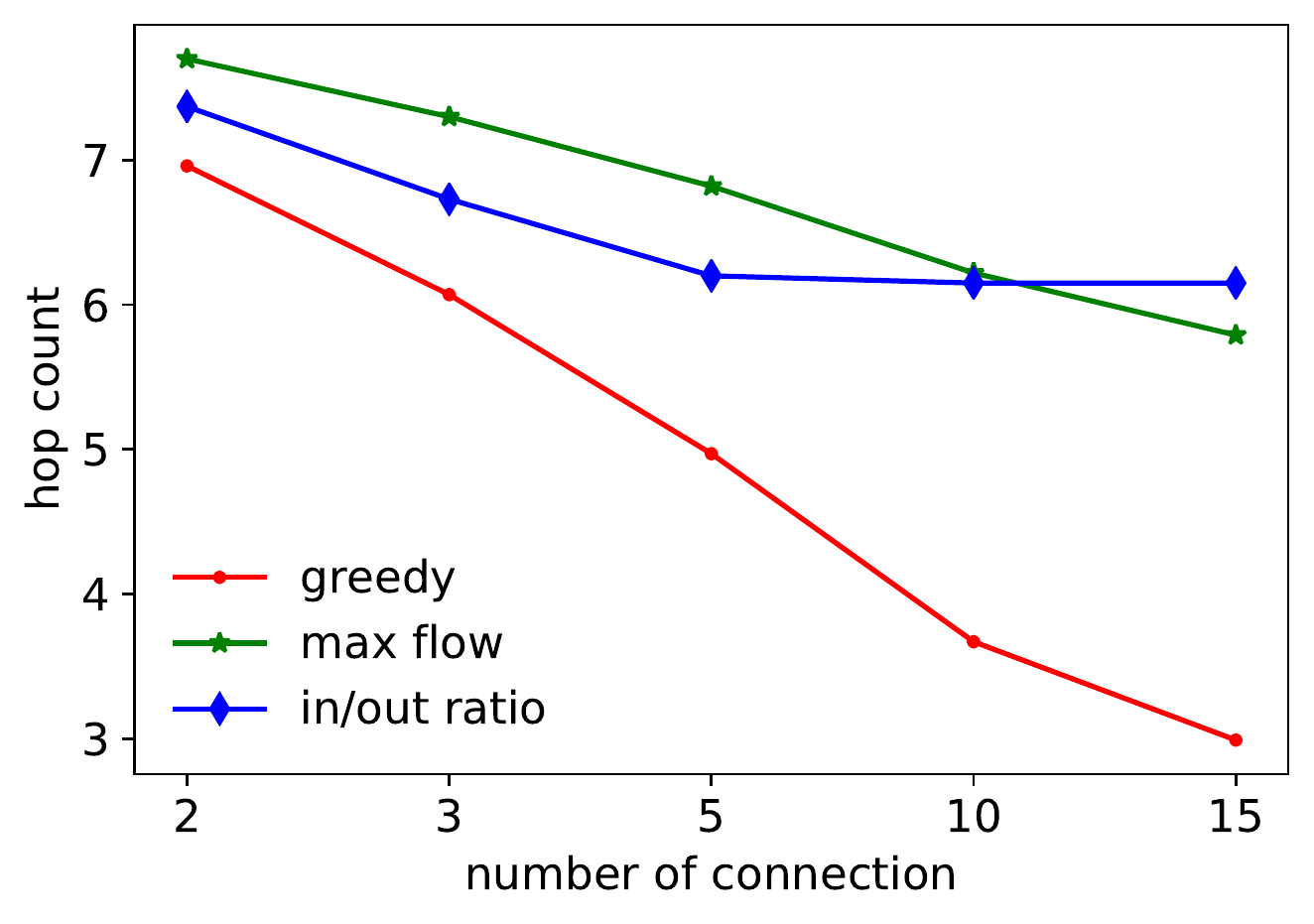} & 
\hspace{-5mm}
\includegraphics[keepaspectratio=true,angle=0,width=58mm]{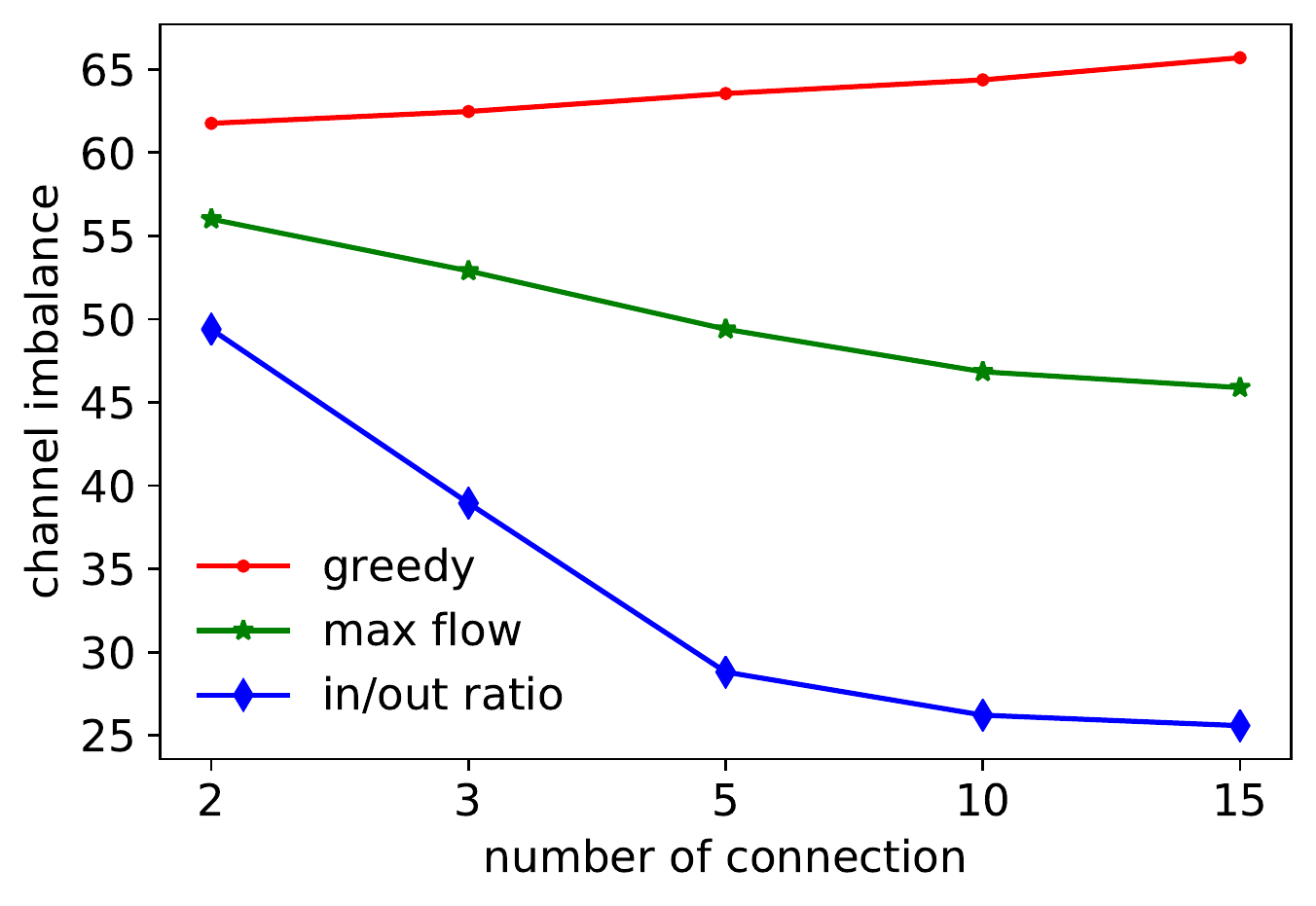} 
\vspace{-2mm}
 \tabularnewline
\footnotesize{ a. Relative Fee} & 
\hspace{-5mm}
\footnotesize{ b. Hop count } &
\hspace{-5mm}
\footnotesize{ c. Channel Imbalance}
\vspace{2mm}
 \tabularnewline

\end{tabular}
\caption{Experiment Results for other factors}
\label{fig:res2}
\vspace{-5mm}
\end{centering}
\end{figure*}

\subsection{Performance Metrics}

We use the following metrics to assess the performance of the proposed approach: 

\begin{itemize}
    \item\textit{Average success rate:} This metric shows the average of the number of payments that could be sent successfully from a source to destination for the whole network.
        
    \item\textit{Fee:} This is the total amount of fees that an end-user has to pay to node owners on the path to destination. Note that in our experiments, the fees are calculated proportional to transfer amount at each hop without a base fee. Obviously, actual fee depends on market value and policy of the underlying payment network.
        
    \item\textit{Network imbalance:} This metric measures the capacity difference between two directions on a channel. We use the average for the network to show the overall imbalance.

    \item\textit{Hop count:} This is the number of hops that an actual payment has traveled when the payment is executed. This could be important if there is a base fee or any delay coming from intermediate nodes.

\end{itemize}

\subsection{Benchmarks}
In order to compare our proposed approach (designated as \textit{in/out ratio} in the graphs), we implemented the following approaches as baselines. In all of these baseline algorithms \enesays{Burasi beni confuse ediyor. Sanki baska algoritmalar da varmis gibi izlenim birakiyor. Anladigim kadariyla bir algoritma ama birkac approach var. Biz ise bir approachu ovuyoruz. Diger approachlar da, baseline olsa da, bizden cikmiyor mu yine?}, we assume that the channel weights (fees) are updated as described in our proposed algorithm. Thus, the only difference is the way that they choose the gateway.

\begin{itemize}
    \item \textit{Greedy approach}:  In greedy approach, the shortest path is computed from connected gateways to the destination in terms of weight, and minimum path among those is selected. The user behaves selfish to minimize the transaction fee. It does not consider either available capacity of the path or the imbalance problem over channels. For instance, if there is one hop path with low capacity, that will be selected to minimize the fee.
    \item \textit{Maximum flow}: This is a general approach adopted by various applications for this type of problems. In this method, instead of choosing the low-cost path in a selfish way, the maximum flow from each gateway to destination is calculated using Ford-Fulkerson flow algorithm \cite{ford2009maximal}. Then, the approach uses the path with highest channel capacity. The idea is leaving available funds in the channels after making a transaction so that upcoming payments will have higher chances to find a proper path. 
    
\end{itemize}

\subsection{Experiment Results}
In the first batch of the experiments, we compare \textit{greedy}, \textit{max-flow} and \textit{in/out ratio} methods when they are using single payments. The success rates are reported in Fig.\ref{fig:res1} while other metrics are reported in Fig.\ref{fig:res2}. These results are the average of various imbalance rate and with payment amounts varying between 25 and 35.

\subsubsection{Success Rate Results}
Looking at the success rate with respect to the number of user connection, we observe that each method is performing better with higher number of connections to gateways as it increases the chances of having a route to any destination in the network. Having higher number of connections means holding more money in the channels for end-users. Based on the results, there is a diminishing return for excessive connections after three. 
The \textit{greedy} approach performs worst in terms of overall performance of the network while it provides cheapest transaction cost for successful payments as will be seen in Fig. \ref{fig:res2}. One reason for lower performance is that it consumes the channel funds selfishly without taking higher balanced routes into consideration. That causes huge skewness in channels as seen in Fig. \ref{fig:res2}c. The \textit{max-flow} method chooses a gateway with higher total capacity to destination to send the payment which yields better performance compared to the first one since utilizing wider paths will leave space for possible upcoming payments that might use the common channels. However, it incurs higher fee per transaction as seen in Fig. \ref{fig:res2}a.  Our proposed approach gives the best success rate among all. This is because, it focuses on keeping the gateways open for end-users. It is easier to find a path in the inner part of the network if the payment can go through source and destination gateways \enesays{inner part neresi oluyor?}. With the current setup, success rate with 5 connections reaches almost 100\% with around 30\% higher fee than the \textit{greedy} method. It also keeps the channels equally balanced in both directions as seen in Fig. \ref{fig:res2}c which helps the network to stay stable. Given that the fees in payment channel networks are much lower compared to blockchain networks, a small increase in the fees for almost guaranteeing all the transactions is a reasonable price. 

We then looked at the impact of payment amount on the success rate as shown in Fig. \ref{fig:res1}b. 
We observe that all the methods are suffering when the amount is increasing because it becomes harder to find a path. However, our approach still significantly outperforms others. The gap even increases with the increased payment amounts. 

Finally, we looked at the imbalance rate which also effects the success rate significantly. Fig.\ref{fig:res1}c shows that after a 40\% imbalance rate, our approach stands out to still maintain high success rate. Again the reason behind this is its ability to look for paths that will minimally impact the channel capacities.  

\begin{figure*}[!htb]
\begin{centering}
\begin{tabular}{ccc}
\includegraphics[keepaspectratio=true,angle=0,width=56mm]{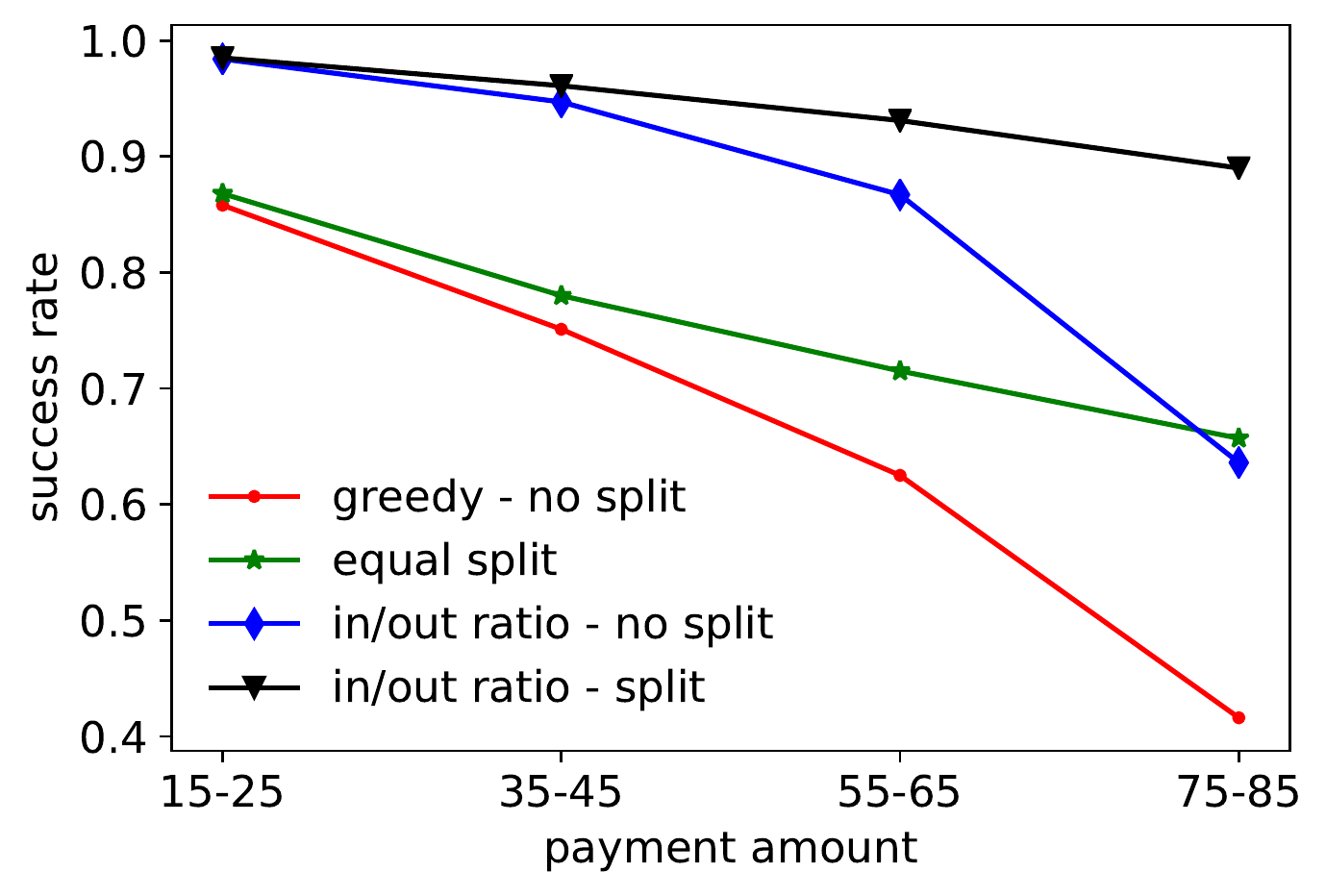} &
\hspace{-5mm}
\includegraphics[keepaspectratio=true,angle=0,width=56mm]{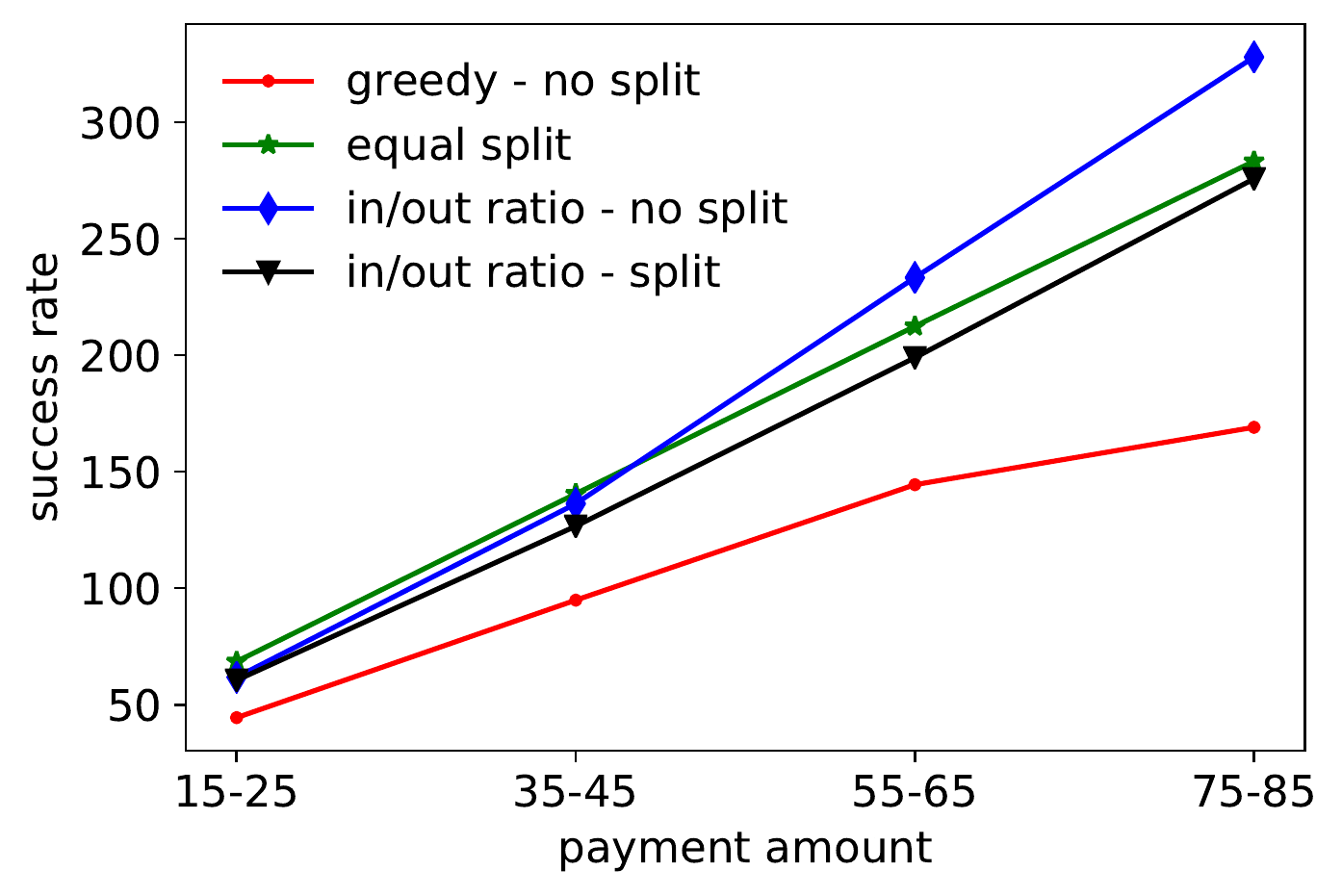} &
\hspace{-5mm}
\includegraphics[keepaspectratio=true,angle=0,width=54mm]{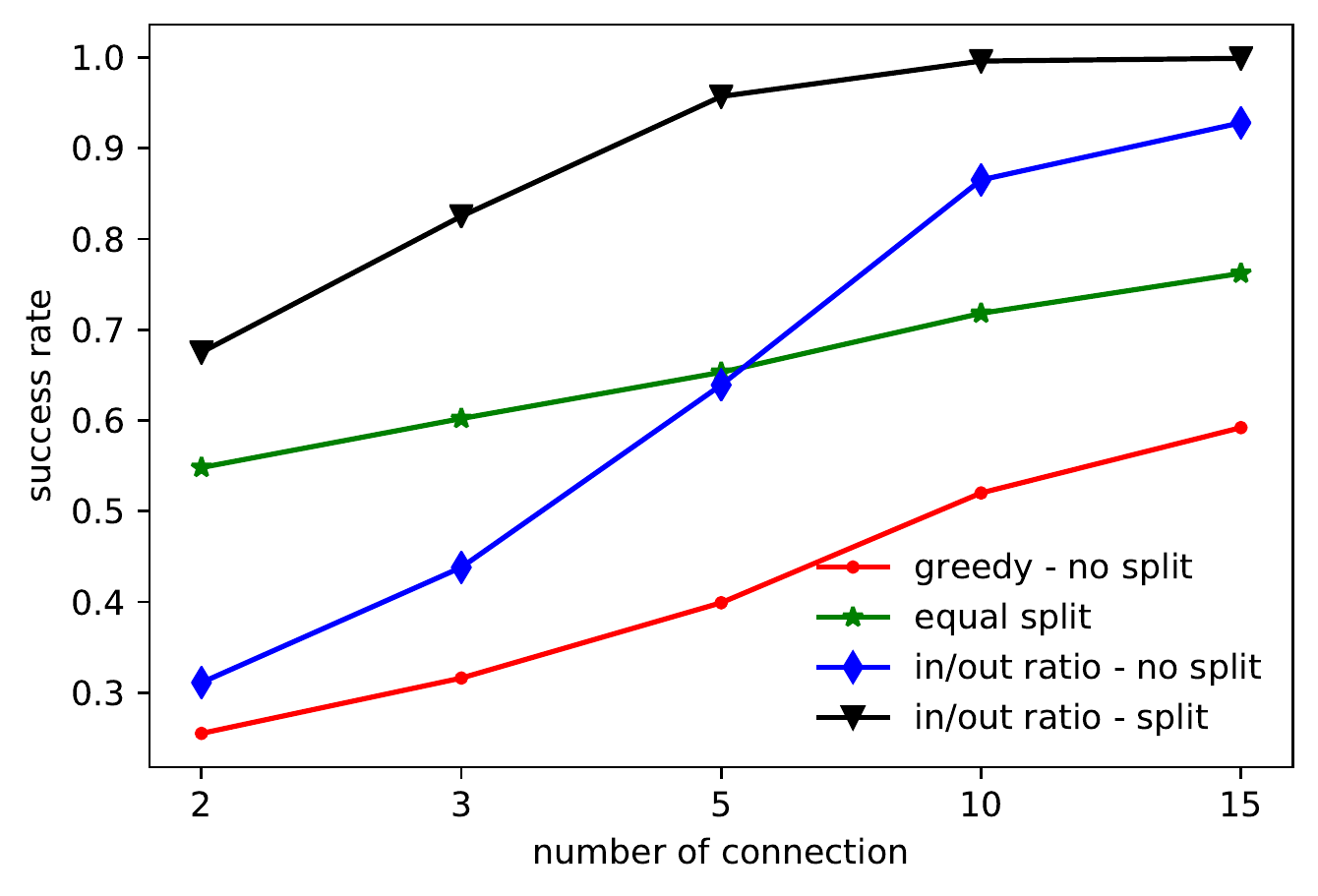} 

\vspace{-2mm}

 \tabularnewline
\footnotesize{ a. Success rate vs. payment amount} &
\hspace{-5mm}
\footnotesize{ b. Success rate vs. fees } & 
\hspace{-5mm}
\footnotesize{ c. Success rate vs. connection count } 
\vspace{2mm}
 
\tabularnewline

\end{tabular}
\caption{Experiment results for split payments.}
\label{fig:res3}
\vspace{-5mm}
\end{centering}
\end{figure*}

\subsubsection{Impact on other Metrics}
Next, we assessed the impact of the proposed approach on the other metrics such as fee, hop count and network imbalance. 

Fig.\ref{fig:res2}a compares the relative transaction fees for the proposed methods. 
Increasing number of connections reduces the fee since there will be shorter paths. The fee in \textit{greedy} method decreases linearly because it always chooses the cheapest route, and extra connections will provide better options in this respect. On the other hand, it stays steady for our approach after a certain point because we try to assign payments based on gateway channel balances. Note that based on the actual fee, the difference between the methods may not be very significant. 

Fig. \ref{fig:res2}b shows number of hops that a payment has to travel to arrive at the destination. The trends are similar to fee results since the fee is directly related to number of hops it has to travel. Average number of hops is also related to degree of the nodes which is set as three in our test network. 

In Fig. \ref{fig:res2}c, we quantify the channel imbalance which basically shows the difference between inbound and outbound capacities in the channels. The difference should be kept lower so that the payments can be transferred from both ways. The \textit{greedy} method makes the network suffer from imbalance problem and causes higher number of failures while our method method relieves it especially after 5 connections. The improvement over geed and even max flow is significant.  


\subsubsection{Results for Split Payments}
In a separate experiment, we present the success rate results of split payment solutions by comparing it with previous(without split) methods for varying payment amounts and connections. Specifically, 
two split methods are implemented: \textit{equal split} that distributes the payment among connected gateways by dividing into equal amounts to assign to gateways, and \textit{ratio proportional split} that divides the payment based on the inbound/outbound ratio. Thus, there are four approaches we compare: 1) greedy approach without splitting (greedy - no split); 2) our approach without splitting (in/out ratio - no split); 3) equal split (equal split); and 4) ratio proportional split (in/out ratio proportional split). 

In Fig. \ref{fig:res3}a, we varied the payment amount from 20 to 80 to observe the success rate of each approach. As can be seen, for smaller amounts, splitting does not have much impact on the results. However, with increasing transaction amounts the difference becomes obvious. For both split methods, the success rates are decreasing linearly while those of single ones are decreasing much faster. Our proposed ratio proportional splitting performs the best. It beats the no-split version of the same approach significantly especially when the payment amount increases.  
In addition, the equal splitting is better than the \textit{greedy} approach with no splitting in any case because it distributes the payments to all connected gateways. 

When we look at the fees shown in Fig. \ref{fig:res3}b, the difference is not significant and the trends are similar. It should be noted that we do not apply any base fee. If there is a base fee in payment channel network design, then splitting might cost more in proportional to the number of splits. In that case, micro payments (splitting payment into tiny amounts) will not be advantageous. 

Fig. \ref{fig:res3}c shows the success rate against the number of connections. We applied a payment amount between 75-85 in this experiment. Again our approach with ratio proportional splitting is the best. We also observe that when there is more than five connections, our approach with no splitting does better than equal splitting. This is because, there is now enough connections to gateways that provides a natural splitting effect.

\section{Conclusion}
In this paper, we presented a smart gateway selection method for end-users in a payment channel networks to send their payments in order to improve the success rate of these payments. We then provided split methods to further increase this success rate. We investigated the effectiveness of the proposed methods by comparing them with two common approaches with extensive experiment analysis. The results indicated the effectiveness of our gateway selection and split methods in achieving high success rates. 

For these results, we observed that keeping the gateways open in terms of channel balance to send and receive payments is an important objective which can be achieved by considering inbound and outbound capacity balance. Split payment is also a effective especially for higher amounts of payments. 

\bibliographystyle{IEEEtran}


\end{document}